\documentclass[prl,twocolumn]{revtex4}
\usepackage{amsmath}
\usepackage{amsfonts}
\usepackage{epsfig}

\begin{document}

\title{Quantum quincunx in cavity quantum electrodynamics} 
\author{Barry C. Sanders}
\affiliation{Department of Physics and Centre for Advanced Computing
  -- Algorithms and Cryptography, \\
        Macquarie University, Sydney, New
  South Wales 2109, Australia} 
\author{Stephen D. Bartlett}
\affiliation{Department of Physics and Centre for Advanced Computing
  -- Algorithms and Cryptography, \\
        Macquarie University, Sydney, New
  South Wales 2109, Australia} 
\author{Ben Tregenna}
\affiliation{Optics Section, Blackett Laboratory, Imperial College
  London, London SW7 2BW, England}
\author{Peter L. Knight}
\affiliation{Optics Section, Blackett Laboratory, Imperial College
  London, London SW7 2BW, England}
\date{4 April 2002}

\begin{abstract}
  We introduce the quantum quincunx, which physically demonstrates the
  quantum walk and is analogous to Galton's quincunx for demonstrating
  the random walk.  In contradistinction to the theoretical studies of
  quantum walks over orthogonal lattice states, we introduce quantum
  walks over \emph{nonorthogonal} lattice states (specifically,
  coherent states on a circle) to demonstrate that the key features of
  a quantum walk are observable albeit for strict parameter ranges.  A
  quantum quincunx may be realized with current cavity quantum
  electrodynamics capabilities, and precise control over decoherence
  in such experiments allows a remarkable \emph{decrease} in the
  position noise, or spread, with increasing decoherence.
\end{abstract}
\maketitle

Galton's quincunx~\cite{Galton1889} is a valuable device for
demonstrating the random walk (RW): gravity draws pellets through a
pyramidal structure of pegs, yielding a binomial distribution.  The RW
is of fundamental importance as the underlying process for dissipation
and fluctuation~\cite{Einstein1905} and as a central concept in
research into computer algorithms, which has motivated research into
the quantum walk (QW)~\cite{adz,abnvw,aakv,mackay} as a quantum
counterpart to the RW.  The QW exhibits surprising features such as a
quadratic enhancement of fluctuations and possible exponential
speedups~\cite{kempe} over the RW.  In addition, the QW could be
useful for benchmarking the performance of certain quantum
devices~\cite{tm}.  Following Galton's classical example, we describe
a cavity quantum electrodynamic (CQED) device that exhibits the QW,
with controllable decoherence that can yield a continuous transition
from the QW to the RW.  Although the (energy-conserving) QW on a
circle has been studied~\cite{abnvw}, and a physical realization in
the context of the ion trap has been introduced~\cite{tm}, ideal
lattice states have always been assumed; however, orthogonal localized
lattice states are not realized physically -- typically they would be
constructed as nonorthogonal Gaussian (e.g., coherent) states.  We
prove here that the QW is viable using such nonorthogonal lattice
states for a restricted range of energies, still exhibiting the
dramatic features characterizing the QW.  Moreover, whereas the
fluctuation--dissipation theorem yields increased fluctuations as
losses increase, this `quantum quincunx' exhibits the counterintuitive
result that fluctuations \emph{decrease} as losses increase.

Microwave CQED provides an excellent technology for realizing the
quantum quincunx.  The combined atom+cavity system can be effectively
isolated from the environment, and decoherence can be controllably
introduced~\cite{Rai97}; furthermore new technologies allow the atom
to be struck by a periodic sequence of off-axis microwave
pulses~\cite{Yam02}.  Whereas the RW utilises a random number (a coin
toss) to determine right or left steps by the `walker', the unitary
evolution of the QW demands a `quantum coin' that is rotated from the
heads ($+$) or tails ($-$) state into an equal superposition of these
states via the unitary Hadamard transformation~\cite{nc}.  In the CQED
realization discussed here, a two-level atom traversing the cavity
serves as the quantum coin, and a periodic sequence of $\pi/2$ pulses
implement the `coin flipping' Hadamard transformations.  Between these
Hadamard transformations, the atom interacts with the initially
coherent cavity field via a Raman transition to effect a conditional
phase shift on the cavity field that depends on the state of the atom.
This QW corresponds to the quantum version of the RW on the
circle~\cite{aakv}.  Whereas both walks exhibit increasing phase
fluctuations with time, the QW spreads quadratically faster than the
RW.

A potential realization of the QW for the ion trap~\cite{tm} has a
comparable mathematical description, but that analysis implicitly assumes
the preparation of harmonic oscillator phase states~\cite{Lou73,Peg97}
of ionic motion (analogous to the cavity field state considered here):
such states would correspond to lattice states on the circle, but
preparation of such states is not feasible.  We consider the field
initially prepared in a coherent state, achievable with existing
technology, and consider the consequences of this initial state.

We begin by introducing the formalism of the QW on the circle embedded
in a harmonic oscillator.  The discrete QW, corresponding to the RW on
the circle, requires a Hilbert space of finite dimension~$d$.  For $\{
|j\rangle, j<d \}$ the harmonic oscillator number states with fewer
than $d$ bosons, we introduce the (finite-dimensional) orthonormal
phase state representation~\cite{Peg97,Buz92}
\begin{equation}
  \label{eq:FinitedPhaseStates}
  |\theta_k= 2\pi k/d\rangle=\frac{1}{\sqrt{d}} \sum_{j=0}^{d-1}
   \exp (\text{i}j\theta_k) |j\rangle\,,\quad k\in\mathbb{Z}_d \, ,
\end{equation}
with the Hilbert space for the walker given by $\mathcal{H}_{d}=
\text{span} \{|\theta_k\rangle, k\in\mathbb{Z}_d\}$.  For $\hat{N}$
the number operator on $\mathcal{H}_{d}$, defined by $\hat{N}|j\rangle
= j|j\rangle$, the rotation operator $R_l = \exp(\text{i}\theta_l
\hat{N})$, $l \in \mathbb{Z}_d$ acts on phase states according to $R_l
|\theta_k \rangle = |\theta_{k+l} \rangle$.  That is, $R_l$ rotates a
phase state by an angle $\theta_l$.  The operator $\hat{N}$ is the
generator of these rotations.

A QW is realized by a sequence of alternating transformations,
beginning with a Hadamard transformation of the two-level system (the
coin) and followed by a conditional rotation of the state of the
walker.  The coin is described by a state in a two-dimensional
internal Hilbert space $\mathcal{H}_2$ with basis states
$|\pm\rangle$.  The Hadamard transformation $\mathbf{H} =
\tfrac{1}{\sqrt 2 } (\begin{smallmatrix} 1&1\\
  1&-1\end{smallmatrix})$ acts only on the internal state of the
coin (i.e., on $\mathcal{H}_2$), and transforms a basis state
$|\pm\rangle$ into the superposition $\tfrac{1}{\sqrt 2} (|+\rangle
\pm |-\rangle)$.  For $\hat{\sigma}_z=\left( \begin{smallmatrix} 1 & 0
    \\ 0 & -1 \end{smallmatrix} \right)$, the conditional rotation
operator
\begin{equation}
  \label{eq:ConditionalRotOperator}
  \mathbf{F} = \exp\bigl( \frac{2\pi\text{i}}{d} \hat{N} \otimes
  \hat{\sigma}_z \bigr)
\end{equation}
rotates the state of the walker by an angle $\pm 2\pi/d$ conditioned
on the coin state $|\pm\rangle$; i.e., $\mathbf{F} (|\theta_k\rangle
\otimes |\pm\rangle) = |\theta_{k\pm 1}\rangle \otimes |\pm\rangle$,
leaving the coin state unchanged.  Thus, beginning with the coin in
the $|+\rangle$ state and the walker in the phase state
$|\theta_0\rangle$, the evolution is described by repeated, and
reversible, application of the unitary operation
$\mathbf{U}=\mathbf{F}\mathbf{H}$.  Thus, the coin and walker degrees
of freedom become entangled.  After $n$ iterations, with the
coin+walker in the state $|\Psi_n\rangle =\mathbf{U}^n
|\theta_0\rangle \otimes |+\rangle$, the probability that the walker
is measured at angle $\theta_k$ is
\begin{equation}
  \label{eq:ProbabilityDistribution}
  P_k = \bigl|\bigl(\langle \theta_k | \otimes \langle +|\bigr)
  |\Psi_n\rangle\bigr|^2 + 
  \bigl|\bigl(\langle \theta_k | \otimes \langle -|\bigr)
  |\Psi_n\rangle\bigr|^2 \, .
\end{equation}
This distribution exhibits the quadratic gain in phase diffusion over
the corresponding RW.

\begin{figure}
  \includegraphics*[height=3in,keepaspectratio]{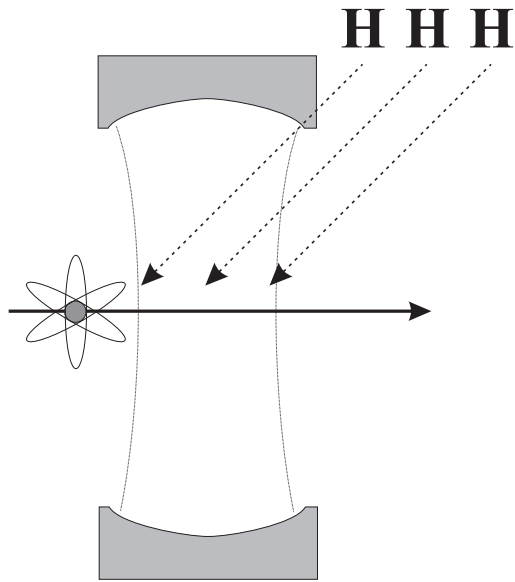}
  \caption{Schematic for the proposed experiment.  A single atom
    traversing through the cavity is subjected
    to periodic Hadamard transformations realized as $\pi/2$ pulses.
    Between these pulses, the cavity field undergoes a phase shift
    conditioned on the atomic state.}
  \label{fig:Schematic}
\end{figure}

We present a scheme to implement a QW on a circle in a microwave
cavity, where the spatial state of the walker is represented by the
state of a single cavity mode, and the state of the coin is
represented by the state of a Rydberg atom passing through the cavity;
a diagram of this scheme is presented in Fig.~\ref{fig:Schematic}.
The field mode in the cavity is described by a harmonic oscillator,
with an infinite-dimensional Hilbert space $\mathcal{H}_{\text{HO}}$.
We wish to use states in this Hilbert space to model a
finite-dimensional QW on a circle with discrete lattice points.  To do
this modelling, we employ a truncated Hilbert space and define the
spatial state of the QW to be given by the projection of the cavity
mode state onto the subspace $\mathcal{H}_{d}$ of states with no more
than $d-1$ photons.  Also, we wish to employ a coherent state as an
initial state for the QW, rather than the unphysical phase state
$|\theta_0\rangle$.  To realize the quantum coin, consider for example
a Rydberg atom with two atomic states $\{ |+\rangle, |-\rangle \}$.

To implement a QW on a circle, we must implement the Hadamard
transformation $\mathbf{H}$ that places the coin in a superposition of
the basis states $|\pm\rangle$, and the conditional rotation
$\mathbf{F}$.  The Hadamard transformation is realized by a $\pi/2$
pulse on the $|+\rangle \to |-\rangle$ transition~\cite{Yam02}.  For
an atom initially in the state $|+\rangle$, this $\pi/2$ pulse
produces the state $\frac{1}{\sqrt{2}}(|+\rangle + |-\rangle)$.  This
Hadamard transformation is assumed to act instantaneously and is
applied with period $\tau$.  To implement the conditional rotation
operator $\mathbf{F}$, we employ the two-level model including
ac-Stark shifts~\cite{mbk,Rai01}.  The atomic levels $|+\rangle$ and
$|-\rangle$ are highly detuned from the cavity field, and the
Hamiltonian for this effect is given by $\hat{H} = \hbar \chi \hat{N}
\otimes \hat{\sigma}_z$.  This Hamiltonian can be used to generate the
conditional rotation operator $\mathbf{F}$ of
Eq.~(\ref{eq:ConditionalRotOperator}) on the subspace $\mathcal{H}_{d}
\subset \mathcal{H}_{\text{HO}}$.  If the atom+cavity evolve according
to this Hamiltonian for time $\tau$ between application of the
Hadamard transformations, the angle of conditional rotation of the
cavity field is given by $\theta = \chi \tau$.

Equivalently, the conditional rotation can be implemented using a
three-level system as in the experiment of Rauschenbeutel \emph{et
  al}~\cite{Rau99}.  Let $|i\rangle$, $|g\rangle$ and $|e\rangle$ be
the states with principal quantum number $n=49,50,51$ respectively.
The state $|g\rangle$ represents the internal basis state $|+\rangle$,
and the state $|i\rangle$ represents the internal basis state
$|-\rangle$.  Employing an off-resonant transition between $|g\rangle$
and $|e\rangle$ (with the state $|i\rangle$ uninvolved), the effective
Hamiltonian is $\hat{H} = \hbar \chi \hat{N} \otimes |g\rangle \langle
g|$.  By moving to a rotating description, this Hamiltonian can effect
the conditional rotation operator $\mathbb{F}$.  The Hadamard
transformation is realized by a $\pi/2$ pulse on the $|g\rangle \to
|i\rangle$ transition.

It is important that the same quantum coin (realized as the Rydberg
atom) is used for each step of the QW, because the atomic state
becomes entangled with the state of the field.  Experimentally, this
constraint requires that the sequence of alternating $\mathbf{H}$ and
$\mathbf{F}$ transformations must be implemented during the passage
time of a single atom.

The standard initial conditions for the QW would be to have the field
(the walker) in the phase state $|\theta_0\rangle$.  Constructing a
field state that projects to this phase state in $\mathcal{H}_{d}$ is
not feasible.  However, it is possible to initiate the cavity in a
coherent state $|\alpha\rangle$, with $\alpha$ real and positive, that
has a well-defined phase relative to the local oscillator used for
homodyne detection.  Let $|\alpha\rangle_d$ be the projection of
$|\alpha\rangle$ onto $\mathcal{H}_d$.  We require that
$|\alpha\rangle_d$ satisfies the overlap condition
\begin{equation}
  \label{eq:Overlap}
  \langle \theta_j |\alpha\rangle_d \simeq \delta_{j0} \,, \quad j \in
  \mathbb{Z}_d \, . 
\end{equation}
For a given dimension $d$, the magnitude of $\alpha$ must be chosen
such that the coherent state $|\alpha\rangle$ has reasonable support
on $\mathcal{H}_d$.  To ensure this support, we employ the condition
$d > \bar{n} + \sqrt{\bar{n}}$, where $\bar{n} = |\alpha|^2$ is the
mean photon number in the coherent state $|\alpha\rangle$.  Also, to
satisfy the overlap condition (\ref{eq:Overlap}), the spacing of the
circular lattice must be sufficiently large.  Defining the standard
quadrature phase space~\cite{Lou87} with $\hat{x} = (\hat{a} +
\hat{a}^\dag)/\sqrt{2}$ and $\hat{p} = (\hat{a} -
\hat{a}^\dag)/\sqrt{2}\text{i}$, a coherent state has a minimum
uncertainty diameter (measured in terms of quadrature standard
deviations) of unity.  For coherent states with mean photon number of
$\bar n$, the circle of radius $\sqrt{\bar{n}}$ can fit approximately
$2\pi\sqrt{\bar{n}}$ distinguishable coherent states.  Thus, we
require that $d < 2\pi\sqrt{\bar{n}}$.  Thus the QW can be performed
only for a range of possibilities for coherent state amplitudes
satisfying $\bar{n} < 28$ and dimension $d < 2\pi\sqrt{\bar{n}}$.

The method of measuring a phase shift of an initial coherent cavity
field using a `homodyning' method~\cite{Rau99} is proposed here to
measure the resulting phase distribution of the cavity field, and thus
analyze the QW.  Once the atom has left the cavity, a coherent local
oscillator field with amplitude $\alpha$ and phase $\varphi$ relative
to the initial field is injected into the cavity, which adds
coherently to the cavity field and gives a resulting amplitude in the
range $0$ to $2\alpha$.  This technique can be utilized to obtain the
probability distribution of the QW as a function of angle for a range
of angles near the initial coherent state.  Obtaining the phase
distribution relies on measuring an ensemble of identical states; it
is key to the successful observation of a QW that the conditions of
the experiment are identical for each run, and that there is no source
of stochasticity that would destroy the quantum interference effects.

We investigate numerically the QW as described above, with $\alpha =
5$ (and thus $\bar{n}=25$) and $d = 31$.  It is assumed that the
Hadamard transformation applied to the atomic states occurs
effectively instantaneously and is independent of the location of the
atom in the cavity.  Cavity losses are simulated via an interaction
between the single-mode cavity field and an external, low temperature
reservoir and are characterized by a loss parameter $g$.  The
atom+cavity thus evolves for a time $\tau$ between Hadamard
transformations by the master equation
\begin{multline}
  \frac{\text{d}}{\text{d}t} \rho(t) = [\chi
  a^{\dagger}a\otimes\sigma_z, \rho(t)] \\ 
  -\frac{g}{2}\left( a^{\dagger}a \rho(t)+\rho(t)a^{\dagger}a +
  2a^{\dagger}\rho(t)a\right),
\end{multline}
where $\chi$ is chosen such that $\chi \tau = 2\pi \text{i}/d$.  Note
that the spatial dependence of $\chi$ on the mode structure of the
cavity can easily be incorporated into the numerical simulations.  A
constant step size $\chi \tau$ could still be maintained with such a
spatial dependence simply by adjusting the frequency of Hadamard
transformations accordingly as the atom traverses the cavity.

\begin{figure}
  \includegraphics*[width=3.25in,keepaspectratio]{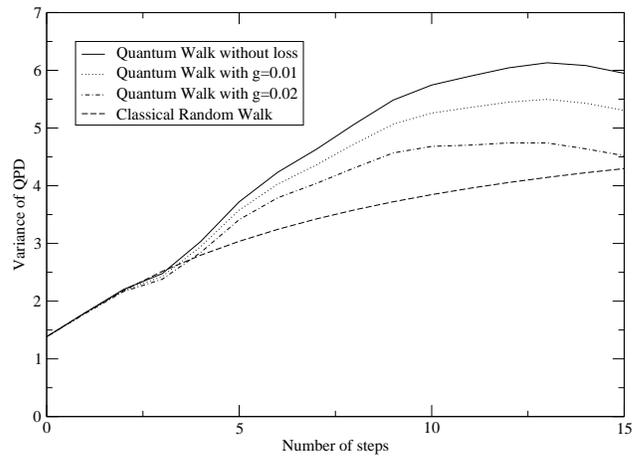}
  \caption{Plot of the quadrature phase variance as a function of the
    number of steps for $\alpha = 5$ and $d=31$.}
  \label{fig:varplot}
\end{figure}

We can simulate the outcome of homodyne measurement and thereby obtain
the resulting quadrature phase distribution (QPD) on the orthogonal
axis to the initial coherent state.  The simulated variance of the QPD
as a function of the number of steps for a lossless cavity is given in
Fig.~\ref{fig:varplot}.
The variance of the QPD of a classical RW for the same values of
$\alpha$ and $d$ obtained by allowing a different atom with a random
atomic state to pass through the cavity during each time step is given
for comparison.  Also shown are the results for the QW in a lossy
cavity with loss term $g=0.01$.  Fig.~\ref{fig:varplot} shows clearly
the quadratic speed-up in phase diffusion given by the QW over the RW
beyond three steps.  This plot also reveals the transition from the QW
to the RW via increasing cavity loss; thus, the addition of
decoherence results in reduced phase fluctuations.  Note that the
variances for the QW and the RW are identical for the first three
steps of the walk (prior to the effects of quantum interference), and
that the initial values of these variances are not zero due to the
width of the initial coherent state.

The QPD approximates the phase distribution for small $\theta$.
Fig.~\ref{fig:varplot} shows that the rate of spreading for the QW is
approximately linear from three to ten steps.  Beyond ten steps, the
rate of spreading decreases as the QPD deviates from the actual phase
distribution.  (For the values of $\alpha$ and $d$ used in the
simulation, the phase distribution of the QW is localised at $\pm \pi$
after 10 steps.)  For this range where the QPD approximates the actual
phase distribution, the system clearly exhibits the
quadratically-enhanced phase fluctuations expected of a QW.

In conclusion, we have shown that a quantum quincunx, which realizes
the QW, can be implemented using existing experimental techniques in a
microwave cavity by taking advantage of a physically realistic,
nonorthogonal basis of coherent states on a circle in phase space.
This quantum quincunx demonstrates the remarkable property that
entanglement between the cavity field and a single atom can lead to
enhanced phase diffusion over an analogous RW, as well as a
controllable transition from the QW to the RW as evidenced by a
decrease in the rate of phase diffusion.  Decreased phase diffusion
resulting from the introduction of decoherence contrasts sharply with
intuition from the fluctuation-dissipation theorem: that the
introduction of loss (decoherence) yields increased noise.  The
quantum quincunx is a remarkable tool to demonstrate a QW, which
provides quadratic or even exponential speedups over the RW, yields a
counterintuitive reduction in phase noise as decoherence increases and
opens the way to new explorations of quantum information theory and
its experimental implementation.

\begin{acknowledgments}
  BCS and SDB acknowledge the support of an Australian Research
  Council Large Grant and a Macquarie University Research Grant.  BT
  and PLK acknowledge the support of the U.\ K.\ Engineering and
  Physical Sciences Research Council.  This project was also funded in
  part by the European Union project QUIPROCONE (IST-1999-29064).  We
  acknowledge helpful discussions with S.\ Haroche, V.\ Kendon, G.\ 
  J.\ Milburn, G.\ Nogues, B.\ C.\ Travaglione and F.\ Yamaguchi.
\end{acknowledgments}


\begin{thebibliography}{99}

\bibitem{Galton1889} F. Galton, \emph{Natural Inheritance}
  (Macmillan, London, 1889).

\bibitem {Einstein1905} A. Einstein, Ann. Phys. (Leipzig)
        \textbf{17}, 549 (1905).

\bibitem{adz} Y. Aharonov \emph{et al}, \pra \textbf{48}, 1687
  (1992).
  
\bibitem{abnvw} A. Ambainis \emph{et al}, ``One-dimensional quantum
  walks,'' in \emph{Proceedings of the 33rd Annual ACM Symposium on Theory
    of Computing} (ACM Press, New York, 2001), p. 37.

\bibitem{aakv} D. Aharonov \emph{et al},
  ``Quantum walks on graphs,'' in \emph{Proceedings of the 33rd Annual
  ACM Symposium on Theory of Computing} (ACM Press, New York, 2001), p. 50.

\bibitem{mackay} T. D. Mackay \emph{et al}, J. Phys. A: Math. Gen.
  \textbf{35}, 2745 (2002).

\bibitem{kempe} J. Kempe, \texttt{quant-ph/0205083} (2002).

\bibitem{tm} B. C. Travaglione and G. J. Milburn, \pra \textbf{65},
  032310 (2002).

\bibitem{Rai97} J. M. Raimond \emph{et al}, \prl \textbf{79}, 1964
  (1997).
  
\bibitem{Yam02} F. Yamaguchi \emph{et al}, \pra \textbf{66}, 010302(R)
  (2002).

\bibitem{nc} M.~A.~Nielsen and I.~L.~Chuang, \emph{Quantum
    Computation and Quantum Information} (Cambridge University Press,
    Cambridge, 2000).
    
\bibitem{Lou73} R. Loudon, \emph{The Quantum Theory of Light}
    (Clarendon, Oxford, 1973).

\bibitem{Peg97} D. T. Pegg and S. M. Barnett, J. Mod. Optics
  \textbf{44}, 225 (1997).

\bibitem {Buz92} V.~Bu\v{z}ek \emph{et al}, \pra \textbf{45}, 8079
  (1992).

\bibitem{mbk} H. Moya-Cessa \emph{et al}, Opt. Comm. \textbf{85},
    267 (1991).

\bibitem{Rai01} J. M. Raimond \emph{et al}, \rmp \textbf{73}, 565 (2001).
  
\bibitem{Rau99} A.~Rauschenbeutel \emph{et al}, \prl \textbf{83}, 5166
  (1999).
  
\bibitem{Lou87} R. Loudon and P. L. Knight, J. Mod. Opt. \textbf{34},
  709 (1987).

\end{thebibliography}
\end{document}